\documentclass[aps,prb,reprint,superscriptaddress, longbibliography]{revtex4-2}

\usepackage{graphicx}
\usepackage{xcolor}
\usepackage{amsmath,amssymb}
\usepackage{braket}

\definecolor{dodgerblue}{HTML}{1E90FF}
\definecolor{lightdodgerblue}{HTML}{4dbff7}
\usepackage[colorlinks=true,citecolor=dodgerblue,linkcolor=dodgerblue,urlcolor=dodgerblue,pdftitle={Dressed symmetry operators}]{hyperref}

\usepackage{amsthm}

\theoremstyle{plain}
\newtheorem{theorem}{Theorem}

\newtheorem{corollary}[theorem]{Corollary}

\theoremstyle{definition}

\theoremstyle{remark}
\newtheorem{remark}[theorem]{Remark}

\begin{document}

\title{Dressed higher-form symmetries in mixed state quantum matter}

\author{Tibor Rakovszky}

\affiliation{Department of Theoretical Physics, Institute of Physics,
Budapest University of Technology and Economics, M\H{u}egyetem rkp. 3., H-1111 Budapest, Hungary}
\affiliation{HUN-REN-BME Quantum Error Correcting Codes and Non-equilibrium Phases Research Group,
Budapest University of Technology and Economics,
M\H{u}egyetem rkp. 3., H-1111 Budapest, Hungary}
\author{Sarang Gopalakrishnan}
\affiliation{Department of Electrical and Computer Engineering,
Princeton University, Princeton, NJ 08544, USA}
\author{Curt von Keyserlingk}
\affiliation{Rudolf Peierls Centre for Theoretical Physics, Physics Department,
Oxford University, Parks Road, Oxford OX1 3PU, United Kingdom}
\begin{abstract}

We describe how to construct emergent strong higher-form symmetries in mixed quantum states that act unitarily on Hilbert space. Our construction uses (quasi-)local recovery channels from quantum error correction to some nearby simple (stabilizer) model. We prove that when the recovery is done in an appropriate (deterministic) manner, the resulting operators continue to be unitary. When applied to operators acting on topologically non-trivial sub-manifolds (loops, surfaces), the resulting unitaries inherit the logical algebra of the fixed point model. More generally, the unitarity of the emergent symmetries allows one to define emergent excitations whose statistics follow those of the fixed point limit.

\end{abstract}

\maketitle

\section{ Introduction} Much of our understanding of phases of matter is organized around the notion of symmetry. This idea has seen renewed enthusiasm in recent years with the realization that even many of the phases that were believed to go beyond the usual notion of spontaneous symmetry breaking (notably, topological order) can in fact be understood in this framework by appropriately broadening the notion of symmetries. Particularly important in this context are so-called \emph{higher-form} symmetries, which act not on the whole system, but on lower-dimensional sub-manifolds~\cite{Gaiottohigherform,mcgreevy2023generalized}. The characteristic (e.g. anyon) excitations of topologically ordered systems arise as the symmetry defects of such higher-form symmetries and their exchange and braiding statistics encode \emph{anomalies} associated to these symmetries~\cite{yoshida_anyon}.

A key difference between usual (0-form) symmetry breaking and its higher form generalizations is the enhanced robustness of the latter. While usual Landau symmetry breaking phases are only well-defined if the system obeys an exact symmetry condition, topological order tends to be robust against \emph{arbitrary} local perturbations~\cite{bravyi2010topological}. This can be understood through the notion of \emph{emergent symmetries}. This notion is made precise in the context of ground state phases of matter through quasi-adiabatic continuation, which constructs a quasi-local unitary operator relating ground states of the perturbed model to those of the original \cite{chen2010local,hastings2005quasiadiabatic}. The same unitary can be used to ``smear out'' the higher-form symmetry operators such that these dressed symmetries will leave the new ground state space invariant. The resulting ``dressed'' symmetries have similar (albeit slightly fattened) supports and have the same algebraic relations as the originals, thus inheriting e.g. the braiding statistics of the excitations they create.

The above picture applies to zero temperature pure ground states. In contrast, mixed states of quantum many-body systems arise naturally once we go from zero to finite temperatures and in more general open quantum systems as a result of (natural or engineered) dissipation. Understanding phases of matter, and particularly topological order, in such mixed quantum many-body states is of much current interest~\cite{ma2025topological,lessa2025strong,tylermixed,AbhinavSohalmixed}. Symmetries again play a crucial role, with the added twist that there are now two distinct kinds: \emph{weak} and \emph{strong}~\cite{albert2014symmetries}. Anomalies of strong higher-form symmetries play a particularly important role~\cite{yoshida_anyon,lessa2025higher,hsin2025higher}. An obvious question in this context concerns the role of emergent symmetries. Suppose a given state has known symmetry generators, and you now perturb it (e.g. by raising its temperature or by applying some local noise to it). Does the resulting state---which is in general mixed---inherit the symmetries of the original? 

There has been important progress on this question~\cite{sangqi_mixedstatephasestop,ellison2024towards,lessa2025higher,sunwoomax} (see Sec.~\ref{SecII} below for a summary), but all existing schemes arguably have downsides: that of Ref. \cite{ellison2024towards} requires extending the Hilbert space while that of Refs. \cite{sangqi_mixedstatephasestop,lessa2025higher} requires that the perturbation is ``locally reversible'' and is mostly relevant to situations that exhibit a mixed anomaly between strong and weak symmetries, rather than anomalies of strong symmetries themselves that are the most relevant to defining quantum topological order~\cite{tylermixed,AbhinavSohalmixed,yoshida_anyon}. While the definition of dressed operators from Refs. \cite{sangqi_mixedstatephasestop,lessa2025higher} still applies to these, it is unclear whether they inherit the algebraic properties (particularly: anomalies) of the `bare' symmetry operators, which are important in defining both the statistics of emergent excitations and the algebra of logical operators. A similar qualification applies to \cite{sunwoomax}. The applications to finite temperature phases in higher dimensions are also less explored.

In this work, we complement these existing results by showing that, under fairly general assumptions, one can explicitly construct emergent higher-form symmetries that, while related to the bare symmetries through quantum channels, nevertheless remain unitary operators on the entire Hilbert space and retain their key algebraic properties, along with their spatial locality. As in Refs. \cite{sangqi_mixedstatephasestop,lessa2025higher}, our definition of emergent strong symmetry generators is based on the idea of \emph{recovery channels} (also referred to as \emph{decoders}) that map states onto a specified code subspace that is defined by the original bare symmetry generators. In particular, we focus on cases when the symmetries are generated by a set of local commuting Pauli stabilizers, which also define a stabilizer code. Our key novel observation is that if the decoder in question has an appropriate property, namely that it is \emph{deterministic}, then the resulting dressed symmetries have a particularly simple structure: they are obtained from the bare symmetries by multiplying (not conjugating!) them with a unitary that commutes with all symmetry operators by construction. 

We show through explicit examples how to construct decoders that are deterministic in the appropriate sense while also being quasi-local, i.e. given by circuits of channels where the size of individual gates and the number of layers are both polylogarithmic in system size (although, as we discuss below, a weaker notion of emergent symmetry goes through even with circuits of constant, rather than polylogarithmic, locality). Whenever both properties hold simultaneously, the notion of anomalies, along with symmetry defects and their braiding/exchange statistics remain well defined and coincide with those of the bare symmetries. This leads us to propose a cleaner symmetry-based definition of topological order in mixed systems akin to the existing definitions of ground state topological phases.

\section{ Quasi-local decoders and emergent symmetries.\label{SecII}}

Our starting point is a mixed quantum state $\rho$ which we take to describe some fixed-point limit of a particular mixed state phase of matter. In particular, we assume that it has a set of \emph{exact} strong higher-form symmetries $W_\gamma^a \rho = \rho$. Here $\gamma$ labels a $p$-dimensional submanifold (making $W_\gamma^a$ a $(d-p)$-form symmetry; e.g. $k=1$ in the case of anyon loops) and $a$ labels different symmetry types (e.g., different anyon species). Given such a state, one can define the type of topological order by examining the anomalies of the strong higher form symmetries, which are encoded in their algebraic relations, a discussed in \cite{tylermixed, AbhinavSohalmixed}. 

Our goal is to extend this discussion to states that are close to $\rho$ in a way we now  make more precise. Consider a ``corrupted'' state $\tilde\rho$. An example to have in mind would be a state obtained by subjecting a 2D toric code state to sufficiently weak local noise. Another would be the thermal Gibbs state of the 4D toric code at sufficiently low temperatures. Common to these examples is the existence of a \emph{recovery channel} $\mathcal{R}$ that undoes the effects of noise and takes $\tilde\rho$ to an appropriate state of the noiseless toric code, $\mathcal{R}[\tilde\rho] = \rho$. It is known \cite{sangqi_mixedstatephasestop,ellison2024towards,lessa2025higher} (see also Refs. \cite{Nahumloopfix1,Nahumloopfix2,liu2026information} for related discussions) that we can use the adjoint of this recovery map to define ``dressed'' operators $\tilde{W}_\gamma^a \equiv \mathcal{R}^\dagger[W_\gamma^a]$.  These generate a strong symmetry of $\tilde\rho$, which follows~\cite{lessa2025higher} from
\begin{align}\label{eq:dressed_symm}
    \langle \tilde{W}_\gamma^a| \tilde \rho\rangle = \langle W_\gamma^a | \mathcal{R}[\tilde\rho]\rangle = \langle W_\gamma^a | \rho\rangle = \text{Tr}(W_\gamma^a \rho) = 1,
\end{align}
where $\langle\cdot|\cdot\rangle$ denotes the Hilbert-Schmidt inner product between operators. 

To make this definition of dressed symmetry operators physically meaningful, $\mathcal{R}$ has to satisfy certain requirements. One such requirement, discussed in \cite{sangqi_mixedstatephasestop} is \emph{locality}. $\mathcal{R}^\dagger$ should be locality-preserving, such that $\tilde W_\gamma^a$ is still meaningfully localized near the submanifold labeled by $\gamma$, at least when this submanifold is sufficiently large. As we will discuss below, in the examples mentioned (as well as many others), $\mathcal{R}$ can be made to act locally on a scale that is logarithmic in the linear size of the system $L$, assuming we are willing to tolerate a small error in the relation $\mathcal{R}[\tilde\rho] = \rho$ that vanishes in the thermodynamic limit. If we only want Eq.~\eqref{eq:dressed_symm} to hold up to a small $O(1)$ error $\varepsilon$, then we will be able to make $\mathcal{R}$ local on a scale that depends on $\varepsilon$ but is independent of $L$. In either case, the geometry of the dressed operators will be maintained on sufficiently large scales.

However, there remains a question about the algebraic properties of these dressed symmetries. In particular, whether they preserve the anomaly structure of the original higher-form symmetries, or, in the language of error correction, whether the symmetry operators associated to non-contractible surfaces satisfy the same logical algebra. In particular, since the dressed operators are obtained by applying a (dual) quantum channel,  \textit{a priori} they have no reason to be unitary operators. Nevertheless, we will now argue that in many cases there are natural choices of $\mathcal{R}$ for which in fact they do remain unitary and also maintain their algebraic relations. It should be mentioned that although $\tilde{W}_\gamma^a$ is not in general unitary, one can show that an extended state $\tilde{\rho}\otimes |0\rangle\langle0|_\mathrm{anc}$ possesses a dressed versions of the original symmetry \cite{sangqi_mixedstatephasestop,ellison2024towards,lessa2025higher} with the correct algebraic properties. In this work, we will attack the problem of trying to define appropriate dressed symmetry operators, with the original algebraic relations, \emph{acting on the original Hilbert space}. 

We will now show how this can be done in a large family of cases of interest. Let us assume that the uncorrupted state $\rho$ is defined in terms of a stabilizer code so that we can partition the Hilbert space into subspaces labeled by the \emph{syndrome} $\mathbf{s}$, i.e. the set of stabilizers being violated. To each subspace, we assign a projector $\Pi_\mathbf{s}$ such that $\sum_\mathbf{s} \Pi_\mathbf{s} = 1\!\!1$. The bare symmetry operator $W_\gamma^a$ is part of the stabilizer group and is thus a Pauli operator that is block diagonal w.r.t. this decomposition: $W_\gamma^a = \sum_{\mathbf{s}} \Pi_\mathbf{s} W_\gamma^a \Pi_\mathbf{s}$. Let us further assume that the Kraus operators of the recovery channel are of the form $K_\mathbf{s} = V_\mathbf{s} \Pi_\mathbf{s}$ where $V_\mathbf{s}$ is a Pauli operator that corrects states within the sector $\Pi_\mathbf{s}$. As $V_\mathbf{s}$ and $W_\gamma^a$ are both Pauli operators, they obey $W_\gamma^a V_\mathbf{s} = V_\mathbf{s} W_\gamma^a e^{\mathrm{i}\theta_{\gamma,\mathbf{s}}^a}$,  (for a qubit sabilizer code $\theta_{\gamma,\mathbf{s}}^a = 0,\pi$ encodes as the two Paulis can only either commute or anti-commute but we also allow more general qudit codes, which can be used to realize any 2D Abelian topological order that admits a gapped boundary~\cite{ellison2022pauli}). From this we find
\begin{align}\label{eq:keyrecoveryeqn}
    \tilde W_\gamma^a = \sum_\mathbf{s} \Pi_\mathbf{s} V_\mathbf{s}^\dagger W_\gamma^a V_\mathbf{s} \Pi_\mathbf{s} = \sum_\mathbf{s} e^{\mathrm{i}\theta_{\gamma,\mathbf{s}}^a} \Pi_\mathbf{s} W_\gamma^a \Pi_\mathbf{s} \equiv U_{\gamma}^a W_\gamma^a,
\end{align}
where $U_{\gamma}^a = \sum_\mathbf{s} e^{\mathrm{i}\theta_{\gamma,\mathbf{s}}^a} \Pi_\mathbf{s}$ is itself a unitary operator. $\tilde W_\gamma^a$ is thus also unitary. Moreover, since $U_{\gamma}^a$ is proportional to the identity within each $\mathbf{s}$-block, it commutes with any of the symmetry operators and thus preserves their algebraic structure. In particular, we may view $U_{\gamma}^a$ is a unitary function of stabilisers operators, and as such commutes with any bare symmetry operator $W_{\gamma'}^{a'}$. It follows immediately that dressed and bare symmetry operators have the same commutation relations $[W^a_\gamma:W^{a'}_{\gamma'}]=[\tilde{W}^a_\gamma:\tilde{W}^{a'}_{\gamma'}]$, where $[A:B]\equiv A^{-1} B^{-1} AB$.

To calculate anomalies, which encode braiding and self statistics, one would also want to consider restricted versions of symmetry operators, which have boundaries where stabilizers are violated (e.g. open line operators creating pairs of anyons). The situation here is slightly more complicated. One can define $U_\gamma^a$ in a similar manner, by keeping track of the phases that arise when commuting $W_\gamma^a$ (with $\gamma$ now being an open submanifold, with boundaries) through $V_{\mathbf s}$. However, the operators $U_\gamma^a$ and $W_{\gamma '}^b$ need no longer commute once at least one of $\gamma,\gamma'$ have boundaries. Nevertheless, as we argue in App.~\ref{app:stats}, the operators $\tilde W_\gamma^a$ maintain the anomalies of their undressed counterparts as long as the distances between the excitations remain larger than the scale on which the decoder $\mathcal{R}$ acts. This ensures that the phases $\theta_{\gamma,\mathbf{s}}^a$ are unaffected by the introduction of these new excitations and therefore drop out of the calculation of the anomaly. Thus, the dressed symmetries have the same anomalies as the bare ones.

One interpretation of this result is as follows. Consider dividing the Hilbert space into different logical sectors based on the eigenvalues of some complete set of logical ``Z'' operators. Within each sector, there is one logical (ground) state as well as many states containing syndromes. We can label a generic basis state as $\ket{\mathbf{z},\mathbf{s}}$, where $\mathbf{z}$ denotes the logical sector and for each syndrome $\mathbf{s}$ we can define a Pauli operator $V_\mathbf{s}$ such that $\ket{\mathbf{z},\mathbf{s}} = V_\mathbf{s} \ket{\mathbf{z},\mathbf{0}}$. Let now $W^x_\gamma$ be a logical ``X''-type operator, defined for some non-contractible $\gamma$, that maps code states from one sector to another: $W_\gamma^x \ket{\mathbf{z},\mathbf{0}} = \ket{\mathbf{z'},\mathbf{0}}$ where $\mathbf{z'}$ is determined by $\mathbf{z}$ and $W_\gamma^x$. Applying the same $W_\gamma^x$ to a generic basis state we get $W_\gamma^x \ket{\mathbf{z},\mathbf{s}} = e^{\mathrm{i}\theta_{\gamma,\mathbf{s}}}\ket{\mathbf{z'},\mathbf{s}}$, picking up an additional phase. What the above definition of $\tilde W_\gamma^x$ achieves is to undo this phase factor so that $\tilde W_\gamma^x \ket{\mathbf{z},\mathbf{s}} = \ket{\mathbf{z'},\mathbf{s}}$ This new, dressed logical operator thus acts the same way as $W^x_\gamma$ on the entire Hilbert space, irrespective of the syndromes. Once these dressed operators are defined for non-contractible sub-manifolds, we can combine them to form operators associated to contractible ones. 

The key ingredient in this discussion was that the decoder $\mathcal{R}$ acts \emph{deterministically}, meaning that for each syndrome $\mathbf{s}$ we associate a unique correction $V_\mathbf{s}$. We could instead have a decoder that chooses randomly between multiple different correction operators, all of which get rid of the same syndrome. In this case, we would no longer end up with a unique phase factor $\theta_{\gamma,\mathbf{s}}$ for a given syndrome and the resulting effect will not generally be unitary. 

The above construction yields a dressed logical that has the correct effect in the whole Hilbert space. This cannot be achieved in a local way due to syndromes that are created by highly non-local errors $V_\mathbf{s}$. However, we do not in fact need to get the phase factors $\theta_{\gamma,\mathbf{s}}$ correctly for \emph{all} syndromes, merely for ones that actually have a significant probability of occurring in $\tilde\rho$. We can thus construct an approximate decoder $\mathcal{R}$ that still manages to approximately clean up the errors in $\tilde\rho$ (and thus restore the exact symmetries) while $\mathcal{R}$ is \emph{both} locality-preserving and deterministic in the above sense. Below, we will discuss how to construct such decoders in a few paradigmatic cases.

The degree of locality we can achieve depends on the precise condition we require from our approximate recovery map. Conceptually, the simplest condition is to ask that $\mathcal{R}[\tilde\rho]$ and $\rho$ are close to each other in trace norm. This means that $\mathcal{R}$ cleans up all syndromes in $\tilde{\rho}$ with high probability and has the advantage that such a recovery can be used to dress any symmetry operators, including those whose size scales with the system size $L$, which is needed to have a non-trivial logical action. However, as the examples below will demonstrate, this typically requires that $\mathcal{R}$ is only quasi-local, in the sense that when we write it as a finite-depth circuit it contains gates whose support is of size $\text{polylog}(L)$. In this case one can make $||\mathcal{R}[\tilde\rho] - \rho||_1$ vanish faster than any polynomial of $1/L$,  so that the symmetries become exact in the thermodynamic limit; however, these symmetries are only well-defined on scales larger than the light-cone of $\mathcal{R}^\dagger$ which is itself diverging (polylogarithmically) with $L$ in this case. This is due to the fact that in a system of linear size $L$, one is likely to find a syndrome of size $O(\log{L})$ \emph{somewhere} in the system. 

On the other hand, to apply the definition of topological order in Refs. \cite{tylermixed,AbhinavSohalmixed}, which involves strong higher-form symmetries and their anomalies, it is sufficient to consider symmetry operators acting on sub-manifolds $\gamma$ whose size remains finite in the thermodynamic limit (but is much larger than the microscopic correlation length). Considering such a $\gamma$, to define the dressed symmetry operator $\tilde W_\gamma^a = \mathcal{R}^\dagger[W_\gamma^a]$, it is sufficient for the decoder $\mathcal{R}$ to clean up those syndromes that would spoil the bare symmetry operator $W_\gamma^a$. As $\gamma$ has finite size, it is unlikely to encounter any of the rare syndromes that cannot be corrected locally, by pairing syndromes within a finite distance. In particular, we can ensure $\text{tr}(\mathcal{R}^\dagger[W_\gamma^a] \tilde\rho) = 1 + \varepsilon$ with a decoder $\mathcal{R}$ that has circuit depth $O(\mathrm{polylog}(1/\varepsilon,|\gamma|))$ that is independent of $L$ for operators with a finite support $|\gamma|$. Below, for simplicity, we stick to the case with $O(\mathrm{polylog}(L))$ depths that ensure vanishing errors in the thermodynamic limit, but adapting our proofs to the finite-depth case should be straightforward. 

\begin{figure*}
  \includegraphics[width=\textwidth]{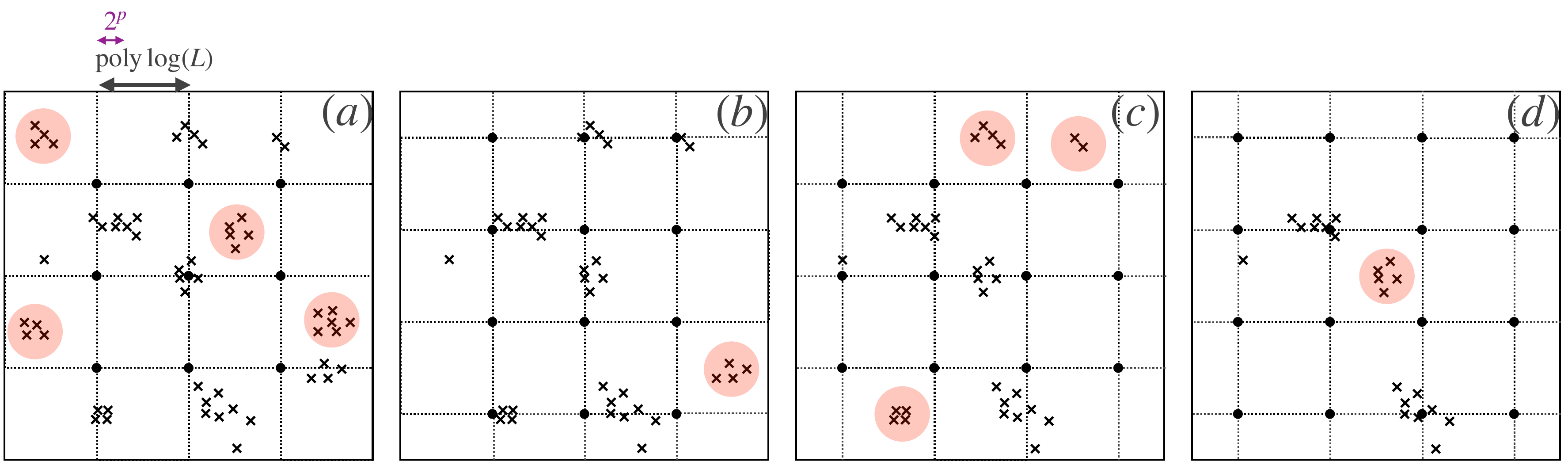}
  \caption{Defect cleaning algorithm at scale $2^p$. The algorithm proceeds to find $2^p$-connected clusters of defects contained entirely within the dashed boxes, at least a distance $2^p$ from the border of the boxes. At each step (a)-(d) the positions of the boxes are shifted by half their widths in the Cartesian directions.\label{defect_fig}}
\end{figure*}

\section{ The decohered toric code model} As a paradigmatic example, let us consider states $\tilde\rho$ obtained by subjecting a pure ground state of the 2D toric code to a weak, finite-depth error channel, such as the ones discussed recently in \cite{bao2023mixed,Fan_mixed_top,wang_fermions,sang_RG_Decoder,chen_grover_sep}. Our goal is to exhibit a decoder that inverts the effects of the noise while being both deterministic (assigning a unique correction Pauli for each syndrome) and quasi-local. We do so by adapting the renormalization group decoder of \cite{BravyiHaahRG}. 

A key observation behind the RG decoder is a decomposition of syndromes into connected clusters. We call a set $k$-connected if any
bipartitioning of it has the resulting two sets within distance at most $k$ of one another. In step $p$ of the RG decoder algorithm, one identifies $2^p$-connected clusters of the syndromes and, if they are are ``anyon neutral'' then corrects them locally. As we show in App.~\ref{app:RG}, for sufficiently small noise (e.g., $\epsilon<10^{-8}$ is a very conservative bound), this procedure terminates before level $p=p_{*}=1+10\log\log L$, and all errors  are removed with probability $1-O(L^{-\infty})$. Another result of this argument is that, with the same high probability, all of the removed $2^p$-connected clusters of errors fit inside boxes with maximum side-length $\ell_*=2^{p_*}=2^{1+10 \log \log L}=O(\mathrm{polylog}(L))$.

To make the RG decoder channel local, we partition the 2D lattice into square boxes of size $\ell_p=6\ell_*$. We consider 4 such partitions, with the boxes shifted relative to each other as in Fig.~\ref{defect_fig}; we call these four partitions $\Lambda_{0,1,2,3}$ respectively. We now break down each step of the RG decoder into 4 sub-steps. In sub-step $0$ we consider only the $2^p$ connected clusters that are enclosed entirely within one of the boxes within $\Lambda_0$ and have a distance $> 2^p$ from the boundary of the enclosing box. For such clusters, we correct them if they are anyon neutral and leave them alone otherwise. We then repeat, with the same $p$, for $2^p$-connected clusters within the boxes corresponding to $\Lambda_1,\Lambda_2$ and finally $\Lambda_3$. We then iterate $p$ until $p=p_*$. All clusters that would have been corrected by the original RG decoder by the $p_*$-th step are corrected by the new procedure;  the key point, elaborated in the App.~\ref{app:RG}, is that for any $p\leq p_*$, all $2^p$-connected clusters lie comfortably inside a box of length $\ell_*$. This box in turn lies entirely within one of the length $6\ell_*$ boxes in one of the partitions $\Lambda_{0,1,2,3}$, and is separated at least  by $2^{p}$ (indeed by $\ell_*$) from the boundary, so will be removed by our modified algorithm. 

\section{ Finite temperature SSB phases}\label{sec:Ising}

\begin{figure*}
  \includegraphics[width=\textwidth]{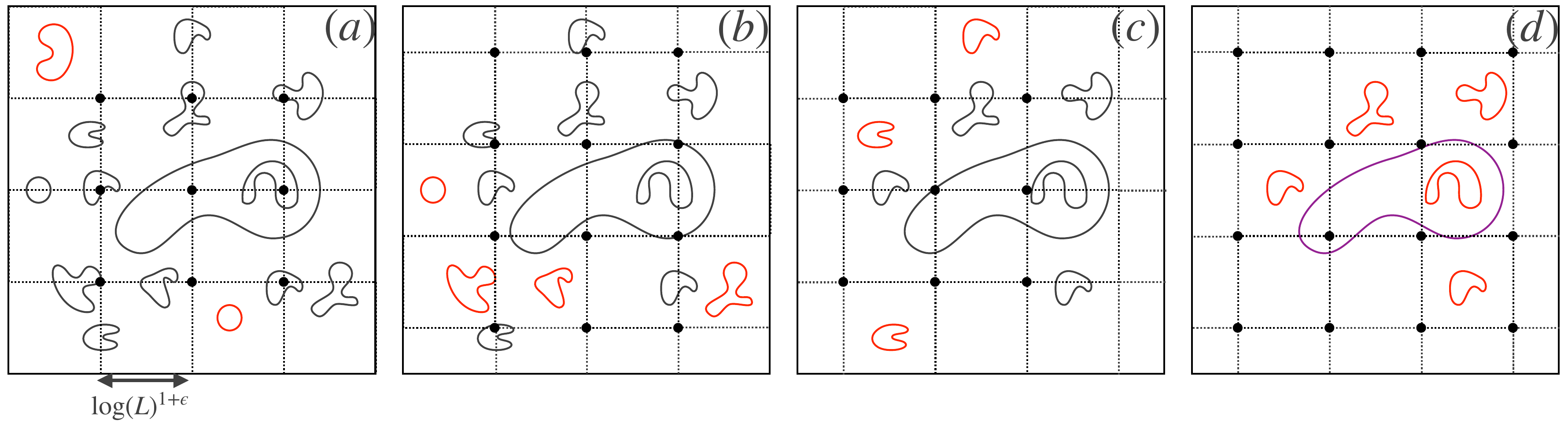}
  \caption{Loop cleaning algorithm for the finite temperatiure Ising model. The algorithm proceeds examines a grid with boxes of size $[\log(L)]^{1+\epsilon}$. Whenever domains are entirely contained in a box, the domain is removed (red). The positions of the boxes are shifted by half a box length in every possible direction, and the process repeated. (There are $2^d$ rounds, where $d$ is the spatial dimension). By the end of the process only loops of length greater than $\frac{1}{2}[\log(L)]^{1+\epsilon}$ (purple) remain.\label{domain_fig}}
\end{figure*}

While the above discussion focused on states with topological order, the techniques generalize to more traditional Landau-Ginszburg orders and furnish a method for defining dressed unitary order parameters away from idealized fixed points. As an example, consider the Ising model in $D=2$. The ground states of this model $H_{\mathrm{Ising}}=-\sum_{\langle ij\rangle} Z_i Z_j$ are stabilizer states defined by $Z_i Z_j =I$ and $\prod_j X_j =\pm I$. In other words, the model has a $\mathbb{Z}_2$ order parameter ($Z^2_i=I$) which exhibits long-range order and is Ising odd---i.e., has an anomaly with the global Ising symmetry. We will show that the finite temperature Ising model, below the ordering transition, has dressed versions of these $Z$ operators, exhibiting long-range order and the same algebraic conditions. Concretely, below the ordering transition temperature $\beta>\beta_c$ the state $\rho_\beta \propto e^{-\beta H_{\mathrm{Ising}}}$  has dressed order parameters obeying $\tilde{Z}_i \tilde{Z}_j\rho_\beta = \rho_\beta +O(L^{-\infty})$ provided $d(i,j)\geq O(\log(L)^{1+\epsilon})$. The dressed order parameters have the additional desirable properties of 1) unitarity, $\tilde{Z}_i^2=1\!\!1$; 2) Ising odd-ness $\{\prod_j X_j,\tilde{Z}_j\}=0$ and by construction, 3) (quasi)-locality---they are smeared out over a diameter $O(\log(L)^{1+\epsilon})$ region. 

To show this, it suffices to construct a deterministic decoder analogous to those discussed for topological phases. In the low-temperature ordered phase,  a domain wall in the Pauli $Z$ basis has a  probability that is exponentially suppressed in length. The probability of having a domain wall of length $\ell$ somewhere in the system is upper bounded by $L^D e^{-\alpha \ell}$ where $\alpha>0$ vanishes as one approaches the critical temperature from below, and $D$ is spatial dimension. As a result, the probability of having a domain wall larger than $\log(L)^{1+\epsilon}$ is $O(L^{-\infty})$ for any $\epsilon>0$. Thus any decoder which removes domain walls up to size  $\log(L)^{1+\epsilon}$ will succeed in decoding the state to a zero temperature/domain-wall free configuration with high probability, i.e.,  $\mathcal{R}(\rho_\beta)=\rho_\infty+O(L^{-\infty})$. 

Fig.~\ref{domain_fig} describes a box decoder which achieves this. The decoder is similar in spirit to that discussed in \cite{3DFTC}. It splits the system into boxes of size $\ell_B=\log(L)^{1+\epsilon}$. At each step, the grid of boxes is  translated as in the figure. The domain walls entirely contained within the boxes are deterministically removed at each stage by flipping the corresponding domain. Once completed, the only remaining domain walls are necessarily at least $\frac{1}{2}\log(L)^{1+\epsilon}$ in extent; but the probability of there being such a loop scales as $L^D e^{-O(\log(L)^{1+\epsilon})}=O(L^{-\infty})$ in the low temperature phase. It is straightforward to then argue that the resulting decoder recovers the zero temperature state with high fidelity $\mathcal{R}(\rho_\beta)=\rho_\infty+O(L^{-\infty})$. Moreover, the decoder is deterministic because at each stage the spin flip to be applies is fully determined by the syndrome (i.e., domain wall) configuration. 

By the same arguments as above $\mathcal{R}^\dagger(Z_i Z_j) \rho_\beta = \rho_\beta+O(L^{-\infty})$. Moreover, provided $\mathrm{dist}(i,j)>O(\log(L)^{1+\epsilon})$ we have $\mathcal{R}^\dagger(Z_i Z_j)= \tilde{Z}_i \tilde{Z}_j$ where $\tilde{Z}_i=\mathcal{R}^\dagger(Z_i)$. Putting these equations together 
\begin{equation}\label{eq:dressedZ}
\tilde{Z}_i \tilde{Z}_j\rho_\beta = \rho_\beta +O(L^{-\infty}). 
 \end{equation}

As above, we can use the deterministic property of the decoder to  show that $\tilde{Z}_i=(-1)^{f(\{Z_j Z_k\})}Z_i$ for an integer function $f$ of stabilizers $Z_i Z_j$. The phase pre-factor $(-1)^{f}$  is the parity number of domain walls that cross site $i$ during the decoding process. This dressed operator has the required properties of 1) unitarity $\tilde{Z}_i^2=I$, 2) Ising odd-ness $\{\prod_j X_j,\tilde{Z}_j\}=0$ and 3) satisfying Eq.~\eqref{eq:dressedZ}.

\section{Higher dimensions and finite temperature topological order} Our results naturally apply to topologically ordered states in finite-temperature equilibrium systems. While such systems are not known to exist in two dimensions, they do exist in higher dimensions. A well-known example if the four-dimensional ``loop-loop'' toric code  \cite{dennis2002topological}. The first example of finite-temperature topological order in three dimensions was constructed in \cite{3DFTC}. 

The key ingredient which makes these various example stable at finite temperature is that the relevant excitations are extended loop-like objects, similar to the domain walls of the 2D Ising model discussed above, rather than the point-like anyons of 2D topological order. This makes a quasi-local correction of syndromes much more straightforward, as was already pointed out in \cite{3DFTC}. In particular, we can provide such a decoder using a slight generalization of the construction for the 2D Ising model above. We cut the system into boxes, performing error correction in each box separately. Due to the loop-like nature of excitations, we need not resort to an RG decoder here (although such a scheme should also work); instead, we can simply consider any loop that is entirely contained within a given box, which can be corrected by a Pauli operator also contained within it. We repeat this for an appropriate number of shifted boxes (consisting of $2^D$ sub-steps in total). After this, \emph{all} loops of linear size smaller than $\ell$ have been corrected, without the need of any further iteration. A Peierls argument, crucial for finite-temperature stability, ensures that the probability of finding a loop of size $\geq \ell$ decays as $L^D e^{-c \ell}$, such that $\ell = O(\log^aL)$ with $a > 1$ suffices for almost-perfect recovery. 

In the case of the 3D fermionic toric code, whose central feature is the fermionic self-stiatistics of point-like emergent excitations, one also wants to use this prescription to define the appropriate dressed line operators that create these local fermions. As discussed in Sec.~\ref{SecII}, the (quasi-)locality of the decoders ensures that such self-statistics are also inherited from the `bare' symmetries, as we argue in App.~\ref{app:stats}. As we also show in App.~\ref{app:stats}, however, for the particular case of the fermionic toric code, there is a more straightforward proof of fermionic self-statistics that does not rely on the aforementioned locality argument. 

In this section, and in Sec.~\ref{sec:Ising} we focused on low temperature thermal states of stabilizer models. However, one can show that in many cases, the Peierls argument survives the addition of small perturbations (with the appropriate 0-form symmetry in the Ising case) to the Hamiltonian, so that loop excitations (defined in terms of the unperturbed Hamiltonian) remain (polylogarithmically) small~\cite{gamarnik2024slow,xun2026local,Yin2026Stability}. Our procedure then defines emergent symmetries for generic states within the same finite-temperature phase sufficiently close to the fixed point. Moreover, similar arguments could be made for states subject to sufficiently weak local noise (again with a symmetry restriction in the Ising case). 


\section{Outlook} In this work, we focused on systems that are in some sense sufficiently close to some commuting stabilizer model. Namely, those that can be corrected back to the stabilizer fixed point using some (quasi-local) deterministic recovery map $\mathcal{R}$. We then showed that the adjoint of the recovery map dresses symmetries in a way that maintains their relevant properties, i.e. both their locality and their symmetry anomalies, the latter of which encode the braiding and self-statistics of excitations as well as the algebra of logical operators. We used explicit constructions to show that an appropriate $\mathcal{R}$ can be constructed in many cases of interest, including the 2D toric code subject to weak noise and various finite temperatures phases of matter based on (perturbations of) stabilizer Hamiltonians. 

For a given recovery map $\mathcal{R}$, the procedure will break down once we go beyond its error threshold, which will likely happen, at least in some cases, before we encounter a genuine intrinsic phase transition (e.g. either a finite-temperature thermodynamic transition or a divergence of the Markov length). This is different from the situation in zero temperature phases, where quasi-adiabatic continuation ensures the existence of dressed symmetry operators everywhere within a gapped phase. It may be possible to construct such local unitary dressed symmetry operators also within the entire mixed state phase by combining a series of smaller recovery maps along the lines of \cite{sangqi_mixedstatephasestop,lessa2025higher}. Proving, or disproving this conjecture is an interesting open problem. 

Another question concerns finite temperature phases perturbed away from the stabilizer fixed point. When quasi-local deterministic decoders can be applied to such models~\cite{gamarnik2024slow,xun2026local,Yin2026Stability}, the existence of unitary dressed symmetries (particularly, of logical operators) should be reflected in the structure of the eigenstates of these Hamiltonians, including the states finite energy-density on which the finite-temperature Gibbs state is concentrated. Working our these implications for highly excited eigenstates is an interesting problem. 

Lastly, in this work, we used stabilizer models as our starting point, which rules out e.g. topological orders with non-Abelian anyons. It is interesting to ask to what extent our results generalize to more general cases, including non-Abelian string-net models. These can still be described using commuting projectors (at least within the ground space) but not generally stabilizers. As a result, the operators that pair up non-abelian syndromes do not have the simple algebraic relations leading to Eq.~\eqref{eq:keyrecoveryeqn}; in particular, the second equality in that expression fails to generalize. It is likely that new ideas are required to extend our results to non-abelian models, if it is indeed possible to do so. 

\begin{acknowledgments}
We thank Meng Cheng, Carolyn Zhang and Samuel Garratt for helpful discussions. S.G. acknowledges support from the
Co-design Center for Quantum Advantage (C2QA) under contract number DE-SC0012704. C.K. is supported by a UKRI FLF MR/Z000297/1. T.R. is supported by the HUN-REN Welcome Home and Foreign Researcher Recruitment Programme 2023, and by the Supported Research Groups
Programme, HUN-REN-BME-BCE Quantum Technology Research Group (TKCS-2024/34). This research was initiated at the Kavli Institute for Theoretical Physics (KITP), supported in part by grant NSF PHY-2309135. ChatGPT-6 Astra was used to improve the readability of the text. 
\end{acknowledgments}

\bibliography{1}

@misc{sunwoomax,
      title={Mixed-state topological order and error-correction thresholds in non-Abelian codes: rigorous results}, 
      author={Sun Woo P. Kim and Max McGinley},
      year={2026},
      eprint={2607.21706},
      archivePrefix={arXiv},
      primaryClass={quant-ph},
      url={https://arxiv.org/abs/2607.21706}, 
}

@article{yoshida_anyon,
  title = {How Much Entanglement Is Needed for Topological Codes and Mixed States with Anomalous Symmetry?},
  author = {Li, Zhi and Lee, Dongjin and Yoshida, Beni},
  journal = {Phys. Rev. X},
  volume = {15},
  issue = {2},
  pages = {021090},
  numpages = {15},
  year = {2025},
  month = {Jun},
  publisher = {American Physical Society},
  doi = {10.1103/pw12-kdjx},
  url = {https://link.aps.org/doi/10.1103/pw12-kdjx}
}

@misc{Nahumloopfix1,
      title={Worldsheet patching, 1-form symmetries, and "Landau-star" phase transitions}, 
      author={Pablo Serna and Andrés M. Somoza and Adam Nahum},
      year={2024},
      eprint={2403.04025},
      archivePrefix={arXiv},
      primaryClass={cond-mat.str-el},
      url={https://arxiv.org/abs/2403.04025}, 
}

@article{Nahumloopfix2,
  title = {Self-Dual Criticality in Three-Dimensional ${\mathbb{Z}}_{2}$ Gauge Theory with Matter},
  author = {Somoza, Andr{\'e}s M. and Serna, Pablo and Nahum, Adam},
  journal = {Phys. Rev. X},
  volume = {11},
  issue = {4},
  pages = {041008},
  numpages = {32},
  year = {2021},
  month = {Oct},
  publisher = {American Physical Society},
  doi = {10.1103/PhysRevX.11.041008},
  url = {https://link.aps.org/doi/10.1103/PhysRevX.11.041008}
}

@article{liu2026information,
  title={Information-theoretic principle of emergent 1-form symmetries},
  author={Liu, Yu-Jie and Xu, Wen-Tao and Pollmann, Frank and Knap, Michael},
  journal={PRX Quantum},
  volume={7},
  number={4},
  pages={043006},
  year={2026},
  publisher={APS}
}

@article{lessa2025higher,
  title={Higher-form anomaly and long-range entanglement of mixed states},
  author={Lessa, Leonardo A and Sang, Shengqi and Lu, Tsung-Cheng and Hsieh, Timothy H and Wang, Chong},
  journal={arXiv preprint arXiv:2503.12792},
  year={2025}
}

@article{ellison2024towards,
  title = {Toward a Classification of Mixed-State Topological Orders in Two Dimensions},
  author = {Ellison, Tyler D. and Cheng, Meng},
  journal = {PRX Quantum},
  volume = {6},
  issue = {1},
  pages = {010315},
  numpages = {44},
  year = {2025},
  month = {Jan},
  publisher = {American Physical Society},
  doi = {10.1103/PRXQuantum.6.010315},
  url = {https://link.aps.org/doi/10.1103/PRXQuantum.6.010315}
}

@article{sangqi_mixedstatephasestop,
  title = {Mixed-State Phases from Local Reversibility},
  author = {Sang, Shengqi and Lessa, Leonardo A. and Mong, Roger S. K. and Grover, Tarun and Wang, Chong and Hsieh, Timothy H.},
  journal = {Phys. Rev. X},
  volume = {16},
  issue = {4},
  pages = {041014},
  numpages = {17},
  year = {2026},
  month = {Oct},
  publisher = {American Physical Society},
  doi = {10.1103/1kr5-jlcb},
  url = {https://link.aps.org/doi/10.1103/1kr5-jlcb}
}

@article{GACS1986,
title = {Reliable computation with cellular automata},
journal = {Journal of Computer and System Sciences},
volume = {32},
number = {1},
pages = {15-78},
year = {1986},
issn = {0022-0000},
doi = {https://doi.org/10.1016/0022-0000(86)90002-4},
url = {https://www.sciencedirect.com/science/article/pii/0022000086900024},
author = {Peter Gács}
}

@article{chen_grover_sep,
  title = {Separability Transitions in Topological States Induced by Local Decoherence},
  author = {Chen, Yu-Hsueh and Grover, Tarun},
  journal = {Phys. Rev. Lett.},
  volume = {132},
  issue = {17},
  pages = {170602},
  numpages = {6},
  year = {2024},
  month = {Apr},
  publisher = {American Physical Society},
  doi = {10.1103/PhysRevLett.132.170602},
  url = {https://link.aps.org/doi/10.1103/PhysRevLett.132.170602}
}

@article{wang_fermions,
  title = {Intrinsic Mixed-State Topological Order},
  author = {Wang, Zijian and Wu, Zhengzhi and Wang, Zhong},
  journal = {PRX Quantum},
  volume = {6},
  issue = {1},
  pages = {010314},
  numpages = {29},
  year = {2025},
  month = {Jan},
  publisher = {American Physical Society},
  doi = {10.1103/PRXQuantum.6.010314},
  url = {https://link.aps.org/doi/10.1103/PRXQuantum.6.010314}
}

@misc{bao2023mixed,
      title={Mixed-state topological order and the errorfield double formulation of decoherence-induced transitions}, 
      author={Yimu Bao and Ruihua Fan and Ashvin Vishwanath and Ehud Altman},
      year={2023},
      eprint={2301.05687},
      archivePrefix={arXiv},
      primaryClass={quant-ph},
      url={https://arxiv.org/abs/2301.05687}, 
}

@article{Fan_mixed_top,
  title = {Diagnostics of Mixed-State Topological Order and Breakdown of Quantum Memory},
  author = {Fan, Ruihua and Bao, Yimu and Altman, Ehud and Vishwanath, Ashvin},
  journal = {PRX Quantum},
  volume = {5},
  issue = {2},
  pages = {020343},
  numpages = {17},
  year = {2024},
  month = {May},
  publisher = {American Physical Society},
  doi = {10.1103/PRXQuantum.5.020343},
  url = {https://link.aps.org/doi/10.1103/PRXQuantum.5.020343}
}

@article{sang_RG_Decoder,
  title = {Mixed-State Quantum Phases: Renormalization and Quantum Error Correction},
  author = {Sang, Shengqi and Zou, Yijian and Hsieh, Timothy H.},
  journal = {Phys. Rev. X},
  volume = {14},
  issue = {3},
  pages = {031044},
  numpages = {24},
  year = {2024},
  month = {Sep},
  publisher = {American Physical Society},
  doi = {10.1103/PhysRevX.14.031044},
  url = {https://link.aps.org/doi/10.1103/PhysRevX.14.031044}
}

@article{BravyiHaahRG,
  title = {Quantum Self-Correction in the 3D Cubic Code Model},
  author = {Bravyi, Sergey and Haah, Jeongwan},
  journal = {Phys. Rev. Lett.},
  volume = {111},
  issue = {20},
  pages = {200501},
  numpages = {5},
  year = {2013},
  month = {Nov},
  publisher = {American Physical Society},
  doi = {10.1103/PhysRevLett.111.200501},
  url = {https://link.aps.org/doi/10.1103/PhysRevLett.111.200501}
}

@article{hastings2005quasiadiabatic,
  title={Quasiadiabatic continuation of quantum states: The stability of topological ground-state degeneracy and emergent gauge invariance},
  author={Hastings, Matthew B and Wen, Xiao-Gang},
  journal={Physical review b},
  volume={72},
  number={4},
  pages={045141},
  year={2005},
  publisher={APS}
}

@article{chen2010local,
  title={Local unitary transformation, long-range quantum entanglement, wave function renormalization, and topological order},
  author={Chen, Xie and Gu, Zheng-Cheng and Wen, Xiao-Gang},
  journal={Physical Review B—Condensed Matter and Materials Physics},
  volume={82},
  number={15},
  pages={155138},
  year={2010},
  publisher={APS}
}

@article{Gaiottohigherform,
    author = "Gaiotto, Davide and Kapustin, Anton and Seiberg, Nathan and Willett, Brian",
    title = "{Generalized Global Symmetries}",
    eprint = "1412.5148",
    archivePrefix = "arXiv",
    primaryClass = "hep-th",
    doi = "10.1007/JHEP02(2015)172",
    journal = "JHEP",
    volume = "02",
    pages = "172",
    year = "2015"
}

@article{AbhinavSohalmixed,
  title = {Noisy Approach to Intrinsically Mixed-State Topological Order},
  author = {Sohal, Ramanjit and Prem, Abhinav},
  journal = {PRX Quantum},
  volume = {6},
  issue = {1},
  pages = {010313},
  numpages = {20},
  year = {2025},
  month = {Jan},
  publisher = {American Physical Society},
  doi = {10.1103/PRXQuantum.6.010313},
  url = {https://link.aps.org/doi/10.1103/PRXQuantum.6.010313}
}

@article{tylermixed,
  title = {Toward a Classification of Mixed-State Topological Orders in Two Dimensions},
  author = {Ellison, Tyler D. and Cheng, Meng},
  journal = {PRX Quantum},
  volume = {6},
  issue = {1},
  pages = {010315},
  numpages = {44},
  year = {2025},
  month = {Jan},
  publisher = {American Physical Society},
  doi = {10.1103/PRXQuantum.6.010315},
  url = {https://link.aps.org/doi/10.1103/PRXQuantum.6.010315}
}

@article{3DFTC,
  title = {Finite-Temperature Quantum Topological Order in Three Dimensions},
  author = {Zhou, Shu-Tong and Cheng, Meng and Rakovszky, Tibor and von Keyserlingk, Curt and Ellison, Tyler D.},
  journal = {Phys. Rev. Lett.},
  volume = {135},
  issue = {4},
  pages = {040402},
  numpages = {7},
  year = {2025},
  month = {Jul},
  publisher = {American Physical Society},
  doi = {10.1103/n9sq-8cxw},
  url = {https://link.aps.org/doi/10.1103/n9sq-8cxw}
}

@article{dennis2002topological,
  title={Topological quantum memory},
  author={Dennis, Eric and Kitaev, Alexei and Landahl, Andrew and Preskill, John},
  journal={Journal of Mathematical Physics},
  volume={43},
  number={9},
  pages={4452--4505},
  year={2002},
  publisher={American Institute of Physics}
}

@article{mcgreevy2023generalized,
  title={Generalized symmetries in condensed matter},
  author={McGreevy, John},
  journal={Annual Review of Condensed Matter Physics},
  volume={14},
  number={1},
  pages={57--82},
  year={2023},
  publisher={Annual Reviews}
}

@article{bravyi2010topological,
  title={Topological quantum order: stability under local perturbations},
  author={Bravyi, Sergey and Hastings, Matthew B and Michalakis, Spyridon},
  journal={Journal of mathematical physics},
  volume={51},
  number={9},
  year={2010},
  publisher={AIP Publishing}
}

@article{ma2025topological,
  title={Topological phases with average symmetries: The decohered, the disordered, and the intrinsic},
  author={Ma, Ruochen and Zhang, Jian-Hao and Bi, Zhen and Cheng, Meng and Wang, Chong},
  journal={Physical Review X},
  volume={15},
  number={2},
  pages={021062},
  year={2025},
  publisher={APS}
}

@article{lessa2025strong,
  title={Strong-to-weak spontaneous symmetry breaking in mixed quantum states},
  author={Lessa, Leonardo A and Ma, Ruochen and Zhang, Jian-Hao and Bi, Zhen and Cheng, Meng and Wang, Chong},
  journal={PRX Quantum},
  volume={6},
  number={1},
  pages={010344},
  year={2025},
  publisher={APS}
}

@article{albert2014symmetries,
  title={Symmetries and conserved quantities in Lindblad master equations},
  author={Albert, Victor V and Jiang, Liang},
  journal={Physical Review A},
  volume={89},
  number={2},
  pages={022118},
  year={2014},
  publisher={APS}
}

@article{hsin2025higher,
  title={Higher-form anomalies imply intrinsic long-range entanglement},
  author={Hsin, Po-Shen and Kobayashi, Ryohei and Prem, Abhinav},
  journal={arXiv preprint arXiv:2504.10569},
  year={2025}
}

@article{gamarnik2024slow,
  title={Slow mixing of quantum Gibbs samplers},
  author={Gamarnik, David and Kiani, Bobak T and Zlokapa, Alexander},
  journal={arXiv preprint arXiv:2411.04300},
  year={2024}
}

@article{xun2026local,
  title={Local-Channel Stability of Classical $\mathbb{Z}_2$ Lattice Gauge Models},
  author={Xun, Guanxi and Cao, Weiguang and Watanabe, Haruki},
  journal={arXiv preprint arXiv:2610.27939},
  year={2026}
}

@unpublished{Yin2026Stability,
  author={Yin, Chao and Li, Yaodong and Rakovszky, Tibor and Khemani, Vedika},
  title  = "Stability of finite temperature quantum phases of matter",
  note = "In preparation",
  year = "2026"
}

@article{ellison2022pauli,
  title={Pauli stabilizer models of twisted quantum doubles},
  author={Ellison, Tyler D and Chen, Yu-An and Dua, Arpit and Shirley, Wilbur and Tantivasadakarn, Nathanan and Williamson, Dominic J},
  journal={PRX Quantum},
  volume={3},
  number={1},
  pages={010353},
  year={2022},
  publisher={APS}
}

\appendix

\section{Braiding and self-statistics} \label{app:stats}

We now discuss how to define anomalies locally, using dressed versions of operators that create excitations and argue that the anomalies of the bare operators are preserved by the dressing, provided that the excitations remain further apart than the length scale associated with the decoder $\mathcal{R}$. We focus on the case of point-like particles for simplicity. 

Consider the operator $W_\gamma^a$ in the stabilizer model, where $\gamma$ is an open line that creates a particle $a$ and its anti-particle $\bar a$ at its two endpoints. Similarly as in the main text, we define the dressed creation operator as $\tilde W_\gamma^a = W_\gamma^a U_\gamma^a$ with $U_\gamma^a = \sum_{\mathbf{s}} e^{i\theta_{\gamma,\mathbf{s}}^a} \Pi_\mathbf{s}$, where $\theta_{\gamma,\mathbf{s}}^a$ is the phase picked up when commuting $W_\gamma^a$ through $V_{\mathbf{s}}$. Note that now $W_\gamma^a$ itself is block off-diagonal in the syndrome sectors $\Pi_{\mathbf{s}}$ and therefore does not commute with $U_\gamma^a$; whether we put $U_\gamma^a$ to the left or right of $W_\gamma^a$ is an arbitrary choice. On the other hand, the dressing $U_a^\gamma$ does commute with $W_{\gamma '}^b$ for a \emph{closed loop} $\gamma '$ for any particle type $b$, a fact we will use below. We now want to argue that the presence of the dressing $U_\gamma^a$ does not change either the braiding or the self-statistics of the particles $a$

\subsection{Braiding statistics} 
\begin{figure}
    \centering
    \includegraphics[width=0.85\linewidth]{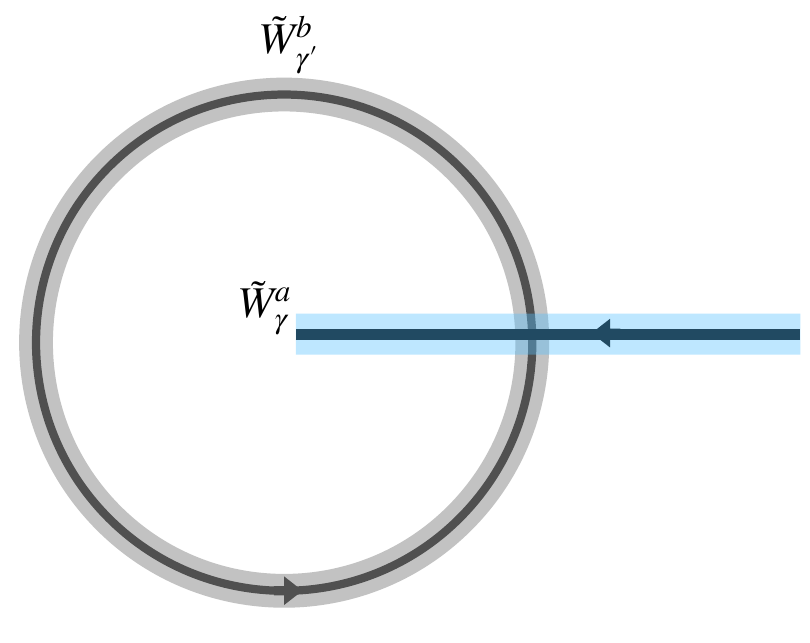}
    \caption{Figure demonstrating the braiding protocol in the text. The solid lines represent the bare operators $W^a_\gamma,W^b_{\gamma'}$, while the shadings represent the dressings  $U^a_\gamma,U^b_{\gamma'}$ due to the decoder. }
    \label{fig:exchange}
\end{figure}
Let us first consider braiding statistics between particles of type $a$ and $b$, which we can extract by considering the commutation between an open string $\tilde W_\gamma^a$ that creates a particle $a$ at some particular location and a closed loop $\tilde W_{\gamma'}^b$ encircling it, which has $a$ on the inside and its anti-particle $\bar a$ on the outside of the loop $\gamma '$ (see Fig.~\ref{fig:exchange}). The braiding statistics in the original stabilizer model (corresponding to anomalies of the strong symmetries of the uncorrupted state $\rho)$ is encoded in the group commutator $[W_\gamma^a : W_{\gamma'}^b] = e^{i \alpha_{ab}}$. We now consider the group commutator of the corresponding dressed operators:

\begin{align}\label{eq:Braiding}
    \tilde W_{\gamma}^a \tilde W_{\gamma'}^b {\tilde W_{\gamma}^{a\dagger}} {\tilde W_{\gamma'}^{b \dagger}} &=  W_{\gamma}^a U_{\gamma}^a W_{\gamma'}^b  U_{\gamma'}^b { U_{\gamma}^{a\dagger}} { W_{\gamma}^{a\dagger}} { U_{\gamma'}^{b\dagger}} { W_{\gamma'}^{b\dagger}} \nonumber \\
    &= W_{\gamma}^a W_{\gamma'}^b  U_{\gamma'}^b { W_{\gamma}^{a^\dagger}} { U_{\gamma'}^{b^\dagger}}{ W_{\gamma'}^{b^\dagger}},
\end{align}
where we used the the fact that $U_\gamma^a$ commutes with both $W_{\gamma '}^b$ and $U_{\gamma'}^b$ and thus cancels with ${U_\gamma^{a\dagger}}$. 

It remains to examine the operator $U_{\gamma'}^b { W_{\gamma}^a}^\dagger { U_{\gamma'}^b}^\dagger$. Imagine acting with this operator on a state with syndrome $\mathbf{s}$. The operator ${ W_{\gamma}^a}^\dagger$ will create a new syndrome configuration which we denote by $\mathbf{s} - \partial\gamma$. We thus have that ${W_{\gamma}^a}^\dagger = \sum_\mathbf{s} \Pi_{\mathbf{s}-\partial\gamma} {W_{\gamma}^a}^\dagger \Pi_{\mathbf{s}}$ and
\begin{align}
    U_{\gamma'}^b { W_{\gamma}^{a\dagger}} { U_{\gamma'}^{b\dagger}} = \sum_\mathbf{s} \Pi_{\mathbf{s}-\partial\gamma} {W_{\gamma}^a}^\dagger \Pi_{\mathbf{s}} e^{i(\theta_{\gamma',\mathbf{s}-\partial\gamma}^b-\theta_{\gamma',\mathbf{s}}^b)}.
\end{align}
To complete the argument, we assume that the decoder $\mathcal{R}$ has some light-cone length scale $\ell$ which is much smaller than the distance of $\gamma'$ from the anyons created by $W_\gamma^a$, $\ell \ll \text{dist}(\gamma',\partial\gamma)$. In calculating the effect of $\mathcal{R}^\dagger$ on $W_{\gamma'}^b$ we could therefore restrict to a region $B_\ell(\gamma')$ of width $\ell$ around $\gamma'$ that does not overlap with $\partial\gamma$. As such, the phase $\theta$ will be insensitive to changing the syndrome configuration at $\partial\gamma$ and we find that
\begin{equation}
    \theta_{\gamma',\mathbf{s}-\partial\gamma}^b = \theta_{\gamma',\mathbf{s}}^b \,\, \forall\mathbf{s} \quad \Rightarrow \quad U_{\gamma'}^b { W_{\gamma}^{a\dagger}} { U_{\gamma'}^{b\dagger}} = { W_{\gamma}^{a\dagger}}.
\end{equation}
Plugging this back into Eq.~\eqref{eq:Braiding} shows that the braiding phase $e^{i\alpha_{ab}}$ remains unchanged by the dressing of the symmetry operators.

The key point in the argument was that the decoder only moves excitations around on a scale $\ell$, while the distance of the new excitations added by $W_\gamma^a$ from $\gamma'$ was much larger than this scale. As a result, the phase $\theta_{\gamma',\mathbf{s}}^b$ was unaffected. Note that in the opposite limit, if the distance was smaller than $\ell$, we would find that the phase would change exactly in such a way as to compensate for the original braiding statistics $\alpha_{ab}$, rendering the commutator trivial. Thus, the braiding statistics remains well defined only for loops and strings on scales much larger than $\ell$.

\subsection{Self-statistics} We now turn to the question of self-statistics. A standard way of defining these is through a ``T-junction'' move, illustrated in Fig.~\ref{fig:tjunct} (left). This involves three open strings, $\gamma_{1,2,3}$ meeting at a point, creating an anyon of type $a$ there. Since all three operators are of the same type, we simplify notation and simply refer to them as $W_{1,2,3}$, respectively. We are then interested in the self-statistics encoded in the relation
\begin{equation}\label{eq:Tjunc}
    W_3 W_2 W_1 W_3^\dagger W_2^\dagger W_1^\dagger = e^{i\alpha_a}.
\end{equation}.

We now want to calculate the same expression for the dressed string operators which yields
\begin{align}\label{eq:Tjunc_dressed}
     &\tilde W_3 \tilde W_2 \tilde W_1 \tilde W_3^\dagger \tilde W_2^\dagger \tilde W_1^\dagger \nonumber \\
     &= W_3 U_3 W_2 U_2 W_1 U_1 U_3^\dagger W_3^\dagger U_2^\dagger W_2^\dagger U_1^\dagger W_1^\dagger.
\end{align}
Overall, the whole expression in both Eq.~\eqref{eq:Tjunc} and~\eqref{eq:Tjunc_dressed} is diagonal in syndromes, since all excitations created by the $W_j$ get destroyed by their adjoints. We will therefore project down the right hand side of Eq~\eqref{eq:Tjunc_dressed} to find 

\begin{align}
     & \Pi_{\mathbf{s}} \tilde W_3 \tilde W_2 \tilde W_1 \tilde W_3^\dagger \tilde W_2^\dagger \tilde W_1^\dagger \Pi_{\mathbf{s}} \nonumber \\
     &= e^{-i\left(\theta_{\gamma_1,\mathbf{s} - \partial\gamma_1} + \theta_{\gamma_2,\mathbf{s}-\partial\gamma_1 - \partial\gamma_2} + \theta_{\gamma_3,\mathbf{s}-\partial\gamma_1-\partial\gamma_2 - \partial\gamma_3}\right)}  \nonumber \\ 
     &\times e^{+i\left( \theta_{\gamma_1,\mathbf{s}-\partial\gamma_1-\partial\gamma_2-\partial\gamma_3} + \theta_{\gamma_2,\mathbf{s}-\partial\gamma_2-\partial\gamma_3} + \theta_{\gamma_3,\mathbf{s}-\partial\gamma_3}\right)}  \nonumber \\ &\times  \Pi_{\mathbf{s}} W_3 W_2 W_1 W_3^\dagger W_2^\dagger W_1^\dagger \Pi_{\mathbf{s}},
\end{align}
where $\partial\gamma_j$ stands for the excitations created by $W_j$. We want to show that the phase factor in the above expression vanishes. This is ensured by satisfying the following three equations:
\begin{align}
    \theta_{\gamma_1,\mathbf{s}-\partial\gamma_1} &= \theta_{\gamma_1,\mathbf{s}-\partial\gamma_1-\partial\gamma_2-\partial\gamma_3} \label{eq:phase_cancel1} \\
    \theta_{\gamma_2,\mathbf{s}-\partial\gamma_1-\partial\gamma_2} &= \theta_{\gamma_2,\mathbf{s}-\partial\gamma_2-\partial\gamma_3} \label{eq:phase_cancel2} \\
    \theta_{\gamma_3,\mathbf{s}-\partial\gamma_1-\partial\gamma_2-\partial\gamma_3} &= \theta_{\gamma_3,\mathbf{s}-\partial\gamma_3}. \label{eq:phase_cancel3}
\end{align}
By assumption, all three operators $W_{1,2,3}$ create the same anyon excitations at the junction. Thus, the two sides of the equations in \eqref{eq:phase_cancel1}-\eqref{eq:phase_cancel3} only differ by excitations that are all far away from location of the string $\gamma_j$. By the same locality argument as in the previous section, the phases are insensitive to adding such far away excitations and thus all three equations are satisfied. Therefore, the dressing does not affect the exchange phases

\begin{equation}\label{eq:Tjunc_final}
    W_3 W_2 W_1 W_3^\dagger W_2^\dagger W_1^\dagger=\tilde W_3 \tilde W_2  \tilde W_1  \tilde W_3^\dagger \tilde W_2^\dagger \tilde W_1^\dagger  = e^{i\alpha_a}.
\end{equation}.
 
\subsection{3D fermionic toric code and 2D Fermion decoherenced toric code} 
\begin{figure*}
    \centering
    \includegraphics[width=0.75\linewidth]{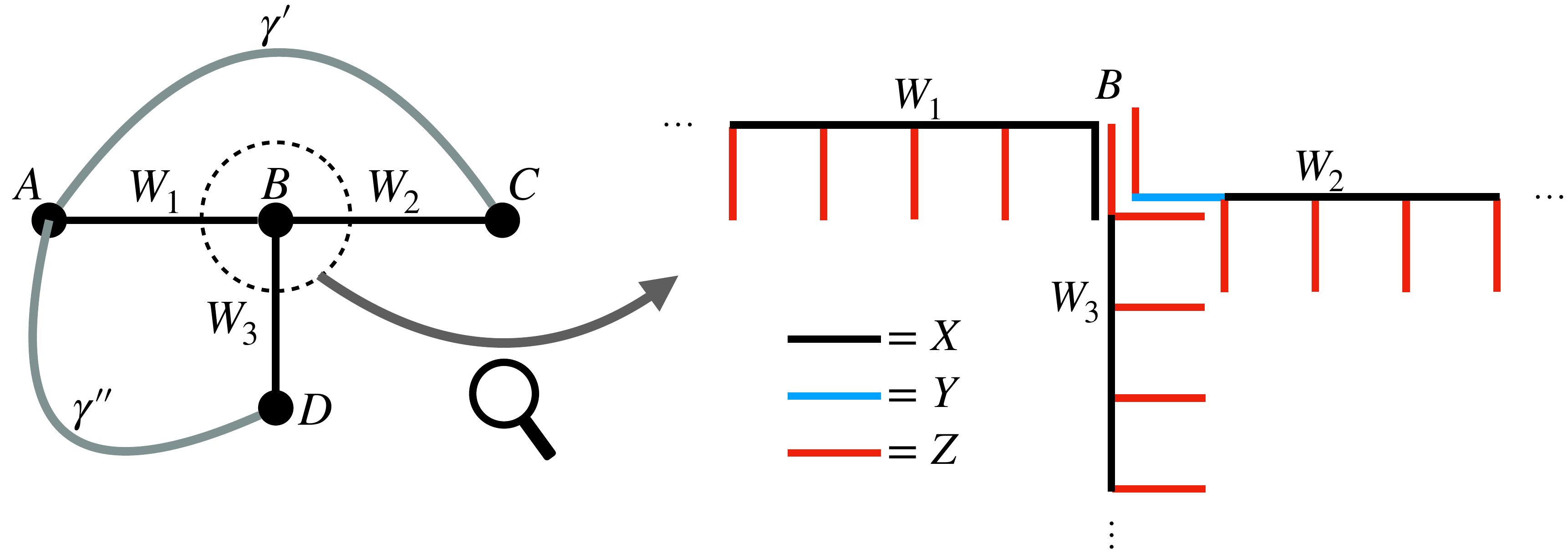}
    \caption{\label{fig:tjunct} The left hand of the figure shows the $T$ junction between three open line operators $W_{1,2,3}$. The line segments $\gamma',\gamma''$ are used later when arguing that anomalous fermionic symmetries imply long-range entanglement App.~\ref{ssec:yoshida_arg}). In the special case of the fermionic decoherend toric code stabilizer state, we show a possible explicit form of these line operators near their junction at $B$ on the right of the figure. }
\end{figure*}

\begin{figure}
    \centering
    \includegraphics[width=0.98\linewidth]{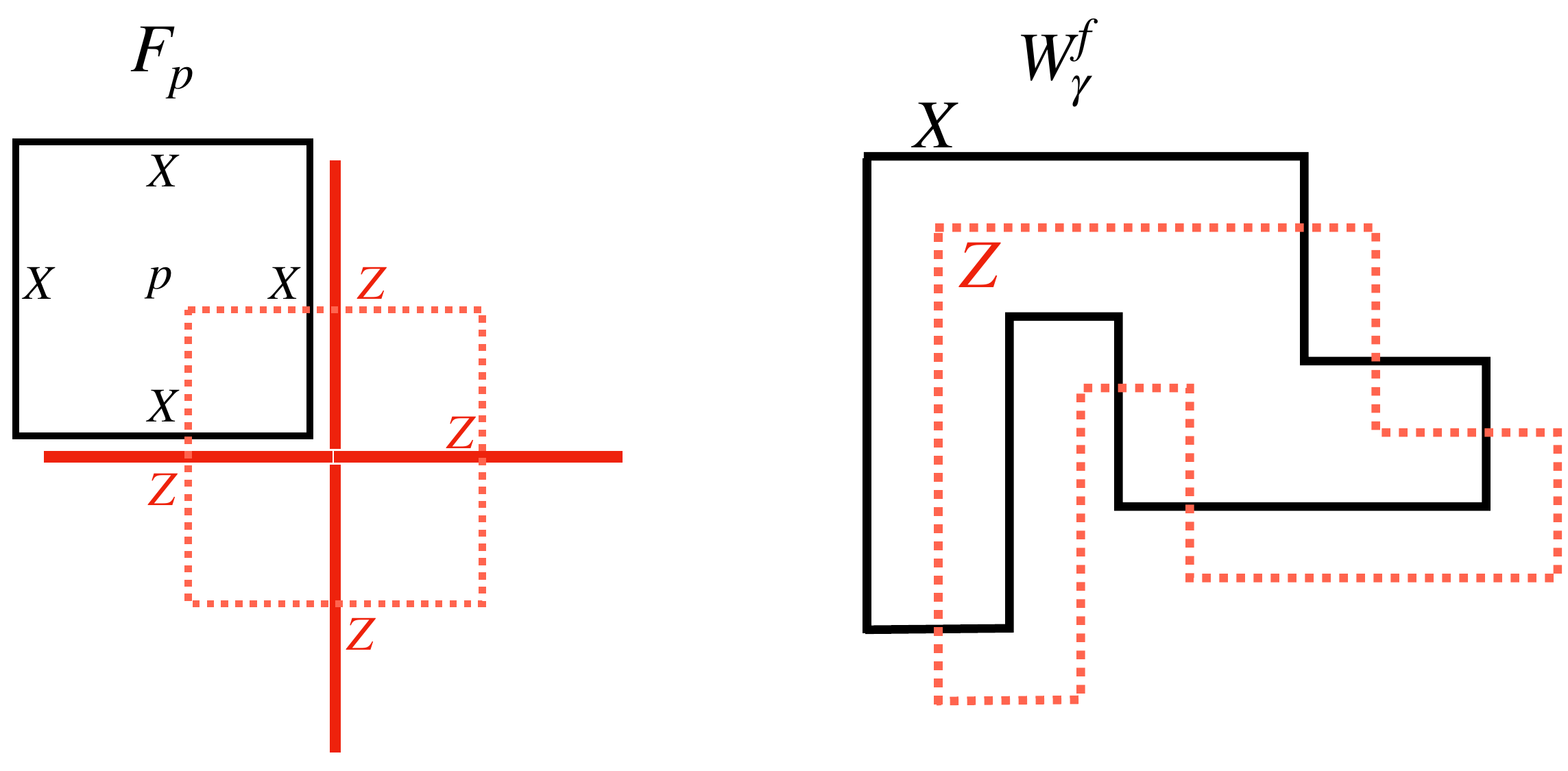}
    \caption{\label{fig:fdtcops} The $F_p$ stabilisers of the fermion decohered toric code consist of a product of a standard toric code plaquette operator with an adjacent vertex operator (left). The plaquette operator is a product of $X$ around a contour on the direct lattice; the associated vertex operator is a product of $Z$'s around  contour on the dual lattice, obtained by shifting the original contour by vector $\frac{1}{2}(1,-1)$. The strong-1 form symmetry generators are obtained similarly (right): A product of $X$'s on a loop on the lattice multiplied by a product of $Z$'s along a $\frac{1}{2}(1,-1)$-shifted contour on the dual lattice.}
    
\end{figure}
It is worthwhile to discuss the paradigmatic examples of the 2D fermion-decohered toric code and the 3D fermionic toric code (at $T>0$) as special cases of the above discussion. In these models, the emergent strong symmetry corresponds to point-like excitations with fermion self-statistics, which is crucial for the non-trivial mixed-state quantum topological order in these states. These examples have some additional structure that allows for a more direct proof, without having to rely on the above argument that was based on locality. Instead, we can argue using the fact that the syndromes in these cases fall into two distinct types and the decoder only needs to clean up one type in order to restore the relevant strong symmetry. This clean up does not touch the fermionic excitations and thus leaves their exchange statistics intact as we now argue.

In the 3D fermionic toric code (see Ref. \cite{3DFTC} for details) the syndromes fall into two categories, corresponding to two types of stabilizers. One type are associated to the square plaquettes of the 3D lattice. These obey local constraints such they have to form closed loops (i.e., every cube of the lattice needs to have an even number of excitations around it), which we refer to as ``vortex loops''. The other type of stabilizers correspond to the vertices of the lattice and these can appear as isolated, point-like quasi particles, created at the two endpoints of a string-like operator $W_\gamma^f$. It is these point-like excitations that have a fermionic self-statistics, which can be obtained by considering T-junction (Fig.~\ref{fig:tjunct}, left). The expression in Eq.~\eqref{eq:Tjunc} evaluated for $W^f$ then gives a fermionic sign $e^{i\alpha_f} = -1$. 

The observation made in Ref. \cite{3DFTC} is that the existence of a strong symmetry under $W_\gamma^f$ is equivalent to the requirement that the system has no vortex loop excitations which can be achieved, at low temperatures, via the kind of quasi-local decoder discussed in the main text that shrinks all small vortex loops to zero. Thus, in this case, the decoding procedure yields a state that is itself a highly mixed state, with a finite density of vertex excitations, but nevertheless has the exact fermionic loop symmetry (up to the $O(L^{-\infty})$ error of the decoder). 

The importance of this for the purposes of the present discussion is that the decoding operations, $V_{\mathbf s}$ in this case depend on only the configuration of the vortex loops and are in fact independent of the configuration of the point-like fermion excitations. Consequently, the phases $\theta^f_{\gamma,\mathbf{s}}$ also depend only on the vortex loop configuration. Considering now the T-junction setup, the operators $W_{\gamma_j}^f$ (where $\gamma_j$ are now three open strings meeting at the junction) do not change the configuration of vortex loops. Therefore all the phases in Eq.~\eqref{eq:Tjunc_dressed} are equal and therefore drop out from the equation automatically.

Likewise, we can consider the fermion decohered toric code. This is a mixed state $\rho_{\mathrm{fdtc }}\propto \prod_p (1+F_p)$ with stabilisers $F_p$ shown in Fig.~\ref{fig:fdtcops}. The state has a fermionic 1-form symmetry: For any set of plaquettes on the lattice note that $[\prod_{p\in S} F_p] \rho_{\mathrm{fdtc }}=\rho_{\mathrm{fdtc }}$. Also note that  the product can be expressed as a line operator $\prod_{p\in S} F_p = W^f_{\partial S}$, so that $\rho_{\mathrm{fdtc }}=W^f_{\partial S} \rho_{\mathrm{fdtc }}$ has a 1-form symmetry. This loop operator is fermionic, in the sense that when it is opened up into a line operator, it creates fermionic quasiparticles, with the associated T-junction exchange statistic; we give some explicit terminations of these line operators on the right hand side of Fig.~\ref{fig:tjunct}. 

Suppose $\rho_{\mathrm{fdtc }}$ is exposed to weak noise. For sufficiently weak noise, we can use the recovery procedure discussed in the main text to define a dressed 1-form symmetry. $F_p$ syndromes are measured and paired up according to, for example, the RG decoder. This gives us a prescription for unitarily dressing the fermionic line operator $\tilde W ^f_\gamma = W ^f_\gamma U^f_\gamma$ where $U^f_\gamma$ is a function purely of $F_p$ stabilisers. Analogous to the 3D FTC, in the 2D FDTC the line operators can be terminated in such a way as to commute with all the defining $F$ stabilsers. Operationally that means that again the phases in Eq.~\eqref{eq:Tjunc_dressed} are equal and therefore drop out from the equation automatically. Equivalently $[W^f_\gamma, U^f_{\gamma'}]=0$ for any pair of lines $\gamma,\gamma'$. 

\subsection{Proving LRE from dressed fermionic symmetry operator\label{ssec:yoshida_arg}}
Ref.~\cite{3DFTC} established that the 3D fermionic toric code remains topologically ordered over a finite temperature window $T<T_c$, in the sense that the density matrix is provably not expressible as an incoherent mixture of short-range entangled (SRE) states. We can use the existence of a dressed symmetry operator, along with the technology in \cite{yoshida_anyon}, to streamline the proof given there. The same mode of proof establishes that the 2D FDTC is topologically ordered below a noise threshold; indeed, we will satisfy ourselves with proving this latter statement, as the 2D setup is slightly simpler. 

We expose the 2D FDTC state to some noise $\tilde\rho = \mathcal{N}\rho_{\mathrm{fdtc}}$. Below a noise threshold, we can use a deterministic recovery procedure (cleaning up $F_p$ syndromes with the RG decoder) to obtain a unitary dressed anomalous 1-form symmetry operator $\tilde W_\gamma^f=W_\gamma^f U_\gamma^f$ for $\tilde \rho$. The RG decoder gives a $U_\gamma^f$ that is at worst $\mathrm{poly}\mathrm \log L$ local. 

The $\tilde W_\gamma^f$ have some desirable properties. If we split a path $\gamma$ into two halves $\gamma=\gamma_1\cup\gamma_2$, and are careful to terminate the line operators in such a way that the endpoints do not violate the $F_p$ stabilizers (as in Fig.~\ref{fig:tjunct}, right), then we have that $W_\gamma^f=W_{\gamma_1}^f W_{\gamma_2}^f$ holds for the bare and the dressed operators. It holds by construction for the bare operators. For the dressed operators we reason as follows. Note that $U^f_\gamma = \sum_{\boldsymbol{s}} \Pi_{\boldsymbol{s}} [W^f_\gamma: V_{\boldsymbol{s}}]$. It follows immediately that  $U^f_\gamma=U^f_{\gamma_1} U^f_{\gamma_2}$. Therefore $\tilde W^f_\gamma= W^f_{\gamma_1} W^f_{\gamma_2} U^f_{\gamma_1} U^f_{\gamma_2}$. But by construction the $U$ and $W$ factors commute, so that
\begin{equation}\label{eq:linehom}
    \tilde W^f_\gamma=W^f_{\gamma_1}U^f_{\gamma_1} W^f_{\gamma_2}  U^f_{\gamma_2} =\tilde W^f_{\gamma_1} \tilde W^f_{\gamma_2}. 
\end{equation}
Now, following \cite{yoshida_anyon}, assume contrary to expectations that 
\begin{equation}\label{eq:sep}
    \tilde \rho \overset{?!}{=}\sum_{\alpha} p_\alpha [\psi_\alpha],
\end{equation}
where $[\psi_\alpha]=|\psi_\alpha\rangle\langle\psi_\alpha|$ is a SRE pure state and $p_\alpha\geq 0$. The approximate 1-form symmetry implies 
\begin{equation}
\mathrm{tr}\left(\frac{1+\tilde W^f_{ABC\gamma'}}{2} \frac{1+\tilde W^f_{ABD\gamma''}}{2}\tilde{\rho}\right)=1-O(L^{-\infty})
\end{equation}
This implies that there is at least one separable state for which 
\begin{equation}\label{eq:deform1}
    \tilde W^f_{ABC\gamma'}|\psi_\alpha\rangle, \tilde W^f_{ABD\gamma''}|\psi_\alpha\rangle = |\psi_\alpha\rangle  + O(L^{-\infty}).
\end{equation}
This readily implies that the fermionic line operators are deformable on this state, in particular
\begin{align}\label{eq:deform}
    \tilde W^f_{ABC}|\psi_\alpha\rangle&=\tilde W^{f\dagger}_{\gamma'} |\psi_\alpha\rangle + O(L^{-\infty})\\
    \tilde W^f_{ABD}|\psi_\alpha\rangle&=\tilde W^{f\dagger}_{\gamma''} |\psi_\alpha\rangle + O(L^{-\infty}).
\end{align}

Assuming that $\tilde{\rho}$ admits a decomposition into an incoherent mixture of SRE states Eq.~\eqref{eq:sep}, we have further shown that at least one of these SRE states must admit deformable dressed fermionic line operators Eq.~\eqref{eq:deform}. 

Therefore, if $\tilde{\rho}$ admits such a decomposition, there must be a pure SRE state, admitting a fermionic symmetry operator such that 1) fermionic line operators can be joined end-to-end Eq.~\eqref{eq:linehom}, and 2) exhibit fermionic anticommutation at a T-junction Eq.~\eqref{eq:Tjunc_final}, and 3) are deformable in the way expected of topological 1-form symmetries (Eq.~\eqref{eq:deform1}). [These three conditions hold provided the separations between $A$, $B$, and $C$, as well as between $\gamma'$ and $B$ and between $\gamma''$ and $B$, exceed the combined light-cone extent of the decoder and the finite-depth circuit used to prepare $\psi_\alpha$ from a product state.]

However, Ref.~\cite{yoshida_anyon} showed that SRE pure states do not admit fermionic line symmetries obeying these three properties. Hence, a pure state possessing our dressed fermionic line symmetries cannot be written as a convex sum of SRE states. 

\section{RG decoder and threshold theorem}\label{app:RG}

We review Bravyi-Haah's RG decoder and their proof that it has a threshold \cite{BravyiHaahRG}.
We then show that a modified `local' version of the RG decoder also
has a threshold. We will need a notion of hierarchies in error sets
originally due to G\'acs \cite{GACS1986}, but our presentation closely follows that
of \cite{BravyiHaahRG}. 

Let $E$ denote the set of locations at which errors occur. Any site
$u\in E$ is a \emph{level}-$0$\emph{ chunk. }A \emph{level}-$1$
\emph{chunk} is a subset $S\subseteq E$ containing two different
sites $u\neq w$, with diameter $d(S)\leq Q/2$. Inductively, a \emph{level}-$n$
\emph{chunk }is a subset $S\subseteq E$ containing two disjoint level-$(n-1)$
chunks such that $d(S)\leq Q^{n}/2$. 
\begin{remark}
    Note that a level-$n$ chunk $S$ has $|S|=2^{n}$. Also note that
any given site might participate in chunks at multiple levels in the
hierarchy.

Denote by $E_{n}$ the union of all level-$n$ chunks. Note that $E_{0}=E$.
Also that $E_{n+1}\subseteq E_{n}$ because any level-$(n+1)$ chunk
is made up entirely of level-$n$ pieces. What about \emph{level}-$1$
chunks? Well, the only way that an error $v$ could fail to be in
$E_{1}$ would be if the distance between $v$ and its complement
in $E$ obey $d(v,E\backslash\{v\}>Q/2$. This is likely for sufficiently
small error rate. So when the error rate is small, $E_{1}$ is empty
because all the noise events are well separated. 
\end{remark}

Define $F_{n}=E_{n}\backslash E_{n+1}$. A site is in this set if
it is in a level-$n$ chunk $S$ which is distance $>Q^{n+1}/2$ from
the nearest disjoint level-$n$ chunk. So regard $F_{n}$ as the set
of dilute errors at level-$n$. Note $E=\bigsqcup_{n\geq0}F_{n}$.
Recall a set is\emph{ $r$-connected} if any bipartitioning of it
has the resulting two sets within distance $\leq r$ of one another.
Note we can write any set as a disjoint union of $r$-connected and
disjoint subsets. 
\begin{theorem}
Let $Q\geq6$. Expressing $F_{n}=\bigsqcup_{\alpha}F_{n,\alpha}$
as a disjoint union of $Q^{n}$-connected sets, we have that the $F_{n,\alpha}$
are far from their complement in $E_{n}$ $d(F_{n,\alpha},E_{n}\backslash F_{n,\alpha})>Q^{n+1}/3$.
In particular $d(F_{n,\alpha},F_{n,\beta\neq\alpha})>Q^{n+1}/3$. 
\end{theorem}

\begin{proof}
Consider $F_{n,\alpha}$. It suffices to show that any error $w\in E_{n}$
more than distance $Q^{n}$ from $F_{n,\alpha}$ is at least distance
$>Q^{n+1}/3$ away. Assume the contrary, i.e., that $Q^{n}<d(w,F_{n,\alpha})<Q^{n+1}/3$.
Consider $v\in F_{n,\alpha}$. We must have $Q^{n}<d(v,w)<Q^{n+1}/3$.
As $v,w\in E_{n}$, they must be a part of level-$n$ chunks $C_{v},C_{w}$
respectively. $C_{v,w}$ have diameters $\leq Q^{n}/2$. So as $Q^{n}<d(v,w)$,
$C_{v,w}$ must be disjoint sets. However, they cannot be too far
from one another as

\begin{align*}
d(C_{v}\cup C_{w})
&\leq d(u,v)+d(C_{v})+d(C_{w})\\
&\leq \frac{Q^{n+1}}{3}+Q^{n}\\
&= Q^{n+1}\left(\frac{1}{3}+\frac{1}{Q}\right)
\leq \frac{Q^{n+1}}{2}.
\end{align*}
The latter equality holds provided $Q\geq6$. But then $C_{v},C_{w}$
are both a part of $E_{n+1}$ which contradicts our assumption.
\end{proof}
\begin{remark}
Note that while $d(F_{n,\alpha},E_{n}\backslash F_{n,\alpha})>Q^{n+1}/3$,
it need not be true replacing $E_{n}\rightarrow E$. That is, there
can be lower level sparse errors that are close to $F_{n,\alpha}$. 
\end{remark}

\subsection{Bounding error sets. }

Work on the $D$-dimensional Euclidean lattice. Let $B_{n}$ be a
fixed cubic box of size $Q^{n}$ and $B_{n}^{+}$ a padded box of
size $3Q^{n}$. Let $E$ be the error set. Consider the probabilities

\begin{align*}
p_{n} & =\mathbb{P}(B_{n}\cap E_{n}\neq\emptyset)\\
\tilde{p}_{n} & =\mathbb{P}(B_{n}^{+}\supset\mathrm{level\text{-}\text{n}\:chunk})\\
q_{n} & =\mathbb{P}(B_{n}^{+}\supset\mathrm{2\,disjoint\,level\text{-}\text{(n-1)}\:chunks})\\
r_{n} & =\mathbb{P}(B_{n}^{+}\supset\mathrm{1\,level\text{-}\text{(n-1)}\:chunk}).
\end{align*}
First note that $p_{n}\leq\tilde{p}_{n}$ as whenever a chunk overlaps
with $B_{n}$ it must be contained in $B_{n}^{+}$. Then $\tilde{p}_{n}\leq q_{n}$
because a level-$n$ chunk consists of two level-$(n-1)$ chunks.
Now $q_{n}$ requires two disjoint level $(n-1)$ chunks to be present
in $B_{n}^{+}$. The BK theorem tells us that $q_{n}\leq r_{n}^{2}$. 

But $B_{n}^{+}$ is a collection of $(3Q)^{D}$ boxes of size $Q^{n-1}$.
So the union bound gives $r_{n}\leq(3Q)^{D}p_{n-1}$. Putting it all
together 
\begin{align*}
p_{n}\leq & (3Q)^{2D}p_{n-1}^{2}
\end{align*}

Iterating gives

\begin{align*}
p_{n}\leq(3Q)^{(2^{2}-1)2D}p_{n-2}^{4} & ,
\end{align*}

and eventually

\begin{align*}
p_{n}\leq(3Q)^{(2^{n}-1)2D}p_{0}^{2^{n}}= & (3Q)^{-2D}(\epsilon(3Q)^{2D})^{2^{n}}.
\end{align*}

Thus the probability that a level $n$ chunk intersects a box $B_{n}$
decays doubly exponentially quickly. Using the union bound, this culminates
in the following theorem.
\begin{theorem}
\label{thm:The-probability-of}The probability of finding a level-$n$
chunk anywhere in the system is bounded above by $L^{2}(3Q)^{-2D}(\epsilon(3Q)^{2D})^{2^{n}}$.
\end{theorem}

\subsection{Proving a threshold theorem for the RG decoder}

The RG decoder works as follows. We are given an initial set of syndromes
$S$. Initialise $p=0$, and iterate until $p=p_{*}$; it turns out
that provided the noise is small enough, the number of RG steps $p_{*}$
can be chosen to be $O(\log\log L)$ and the RG decoder will succeed
with a probability exceeding $1-O(L^{-\infty})$. In each step of
decoding, identify the $2^{p}$-connected clusters of the syndrome
$S$, and bound them with minimal boxes. For any such connected clusters
that are anyon neutral, apply a suitable Pauli string to the 1-neighbourhood
of the corresponding bounding box to eliminate the syndromes. Update
$S$, and increment $p\rightarrow p+1$. 
\begin{theorem}
\label{thm:fullcorrection}For $Q\geq18$, if the chunk decomposition
of $E$ terminates at the $m$-th level, and $Q^{m+1}<L/2$ then $E$
is fully corrected by the $RG$ decoder. 
\end{theorem}

\begin{proof}
Let $E=\cup_{j=0}^{m}F_{j}$ be the locations of Pauli-errors. Recall
that $F_{j}$ have $Q^{j}$-connected components which are at least
$Q^{j+1}/3$ separated. Label these $F_{j,\alpha}$; denote the 1-neighbourhood
of $F_{j,\alpha}$'s minimal bounding by $B_{j,\alpha}$. Then by
the nature of these connected components above, we have

\begin{align*}
d(B_{j\alpha}) & \leq Q^{j}+2,\\
d(B_{j\alpha},B_{k\beta}) & \geq\frac{1}{3}Q^{\min(j,k)+1}-1.
\end{align*}
The first condition comes from the $Q^{j}$-connectedness of $F_{j\alpha}$.
The second comes from $d(F_{j\alpha},E_{j}\backslash F_{j\alpha})>Q^{j+1}/3$.
By design, the errors on each $B_{j\alpha}$ are charge neutral; because
$F_{j\alpha}$ contains an isolated collection of errors (not syndromes).
Thus, all of the syndromes in any $B_{j\alpha}$ can be eliminated
by acting solely on $B_{j\alpha}$. 

Note that the minimal separation between any two $B$-sets is $\frac{Q}{3}-1\geq5$.
Recall too that all syndromes are contained inside of the union $\cup_{j=0,\alpha}^{m}B_{j,\alpha}$. 

At $p=0$, the algorithm finds the 1-connected clusters of syndromes.
Each $1$-connected component must be entirely contained inside of
some $B_{j\alpha}$ because any $B_{j\alpha}$ are at least $5$-separated.
Get rid of the resulting 1-clusters that are charge neutral. After
this correction, all $B_{j\alpha}$ must still be neutral because
they started as such. 

At $p=1$, the algorithm finds 2-connected clusters of errors. Once
again, as all $B_{j\alpha}$ are more than 2-separated, each such
2-connected cluster lies entirely within some $B_{j\alpha}$. 

At $p=2$, the algorithm finds 4-connected clusters of errors. At
this point, the only errors remaining in a $B_{0\alpha}$ set are
part of a 4-connected neutral cluster entirely contined in said $B_{0\alpha}$.
That is because any such errors are 5-separated from errors outisde
of $B_{0\alpha}$, and all errors in $B_{0\alpha}$ are within 4-sites
of one another (because $d(B_{0\alpha})\leq3$). The algorithm thus
elminates the $B_{0\alpha}$ sets. It also may eliminate 4-connected
subsets of the $B_{j>0,\alpha}$. 

We are thus at a point $p=2$ where all syndromes contained in $B_{j\alpha}$
with $Q^{j}+2\leq2^{p=2}$ have been eliminated and all remaining
$B$ sets are anyon neutral. We will show inductively that this relation
continues to hold as we iterate $p\geq2$.

Suppose that all syndromes contained in $B_{j\alpha}$ with $Q^{j}+2\leq2^{p}$
have been eliminated. Denote by $j_{*}$ the maximum such $j$ so
that $Q^{j_{*}}+2\leq2^{p}<Q^{j_{*}+1}+2$. Iterate $p$. We now find
$2^{p+1}$-connected clusters of syndromes. These are entirely contained
within remaining $B$ sets because for $j,k\geq j_{*}+1$,
\begin{align*}
d(B_{j\alpha},B_{k\beta}) & \geq\frac{1}{3}Q^{j_{*}+2}-1.\\
 & \geq\frac{Q}{3}(2^{p}-1)-1\\
 & \geq2^{p+1}+2^{p+1}-5>2^{p+1}.
\end{align*}
The algorithm will eliminate such a cluster if it is anyon neutral.
But as the cluster is entirely contained in a $B_{j\alpha}$, it leaves
the anyon charge of $B_{j\alpha}$ unchanged. To complete the induction,
we need to show that once $p$ becomes large enough, the $B_{j_{*}+1,\alpha}$
sets are removed. So suppose that $2^{p}<Q^{j_{*}+1}+2\leq2^{p+1}$
. At stage $p+1$, the algorithm finds $2^{p+1}$-connected clusters.
We know that if any such cluster has support in $B_{j_{*}+1,\alpha}$
it must be entirely contained in that set from the working above.
On the other hand, we know that every error in $B_{j_{*}+1,\alpha}$
is part of the cluster because $d(B_{j_{*}+1})\leq Q^{j_{*}+1}+2\leq2^{p+1}$.
We also know the cluster is anyon neutral by hypothesis. Thus every
$B_{j_{*}+1,\alpha}$, and thus all errors in the $(j_{*}+1)$-th
level of the chunk decomposition, are eliminated by the end of the
$(p+1)$-step where $Q^{j_{*}+1}+2\leq2^{p+1}$. 
\end{proof}
\begin{remark}
The intuition behind this construction is that we erase $B_{j}$ as
soon as $p$ obeys $2^{p}\geq Q^{j}+2>2^{p-1}$. However, the $B_{j}$
sets are widely separated compared to their size ($Q^{j+1}$ vs $Q^{j}$),
and this ensures that the $2^{p}$-connected sets lie entirely in
the $B$'s. Any remaining errors in a $B_{j}$ set are $2^{p}$-connected
and must be removed at this point because $d(B_{j})\leq Q^{j}+2\leq2^{p}$
(charge neutrality is guaranteed by construction for the $B$). 
\end{remark}

\subsection{Application to decoding the toric code in 2D}

Subject a 2D toric code ground state to Pauli dephasing errors, with
local error probability $\epsilon$. Take $\epsilon$ to be small
(say $\epsilon<\frac{1}{2}(3Q)^{-4}$). In this case, \ref{thm:The-probability-of}
implies that the probability of finding a level-$m$ chunk anywhere
in the system is bounded above by $L^{2}p_{m}\leq L^{2}2^{-2^{m}}$.
If we set $m=2\log\log L$ (logs are base $2$), then this probability
is bounded by $L^{2}e^{-\log(L)^{2}}=O(L^{-\infty})$ which decays
superpolynomially in system size. Thus the chunk decomposition truncates
by $m=2\log\log L$ with high probability $1-O(L^{-\infty})$. What
are the consequences for the RG algorithm?

For the purposes of specifying a sufficiently large $p_{*}$, it will
be convenient to take $Q=2^{5}$ from this point onwards. Recall that
the $m$-th level of the chunk decomposition, $F_{m}$, is eliminated
once $Q^{m}+2\leq2^{p}$. In the present case, this is guaranteed
if $p=5m+1$. This leads to the corollary:
\begin{corollary}
For $\epsilon<\frac{1}{2}(3Q)^{-4}$ and $Q=2^{5}$, the RG decoder
will succeed if terminated at $p_{*}=10\log\log L+1$, with a probability
exceeding $1-O(L^{-\infty})$. 
\end{corollary}

We now make a local version of the RG decoder. Note that the $B_{j\alpha}$
in the chunk decomposition have maximum diameter $Q^{m_{*}}+2\leq2^{p_{*}}=2\log(L)^{10}=:\ell_{*}$
with high probability. Let $P_{ab}$ denote the set of squares of
length $2R$ whose origin (lower left corner) lies at a point of form
$(aR+2Rm,bR+2Rn),m,n\in\mathbb{Z}$ where $a,b=0,1$. We pick the
square lengths to be larger than the extent of the likely $B_{j\alpha}$,
e.g., $2R=6\ell_{*}$. With this choice, any $B_{j\alpha}$ that is
likely to occur will lie comfortably in at least one square in the
sets $P_{ab}$, so as to be at least a distance $\ell_{*}$ from the
boundary of said square.

The algorithm proceeds as follows. We iterate on $p$ as in Haah-Bravyi,
terminating at $p_{*}=10\log\log L+1$. At general level $p$, we
first identify all $2^{p}$-connected clusters of errors in each square
in $P_{00}$. But we ignore $2^{p}$-connected clusters that are within
distance $2^{p}$ of the boundary of squares in $P_{00}$. We eliminate
the neutral $2^{p}$-connected clusters obeying this condition as
before. We repeat the same procedure for $P_{01},P_{10},P_{11}$.
By the end of this procedure, we have clearly removed a subset of
the $2^{p}$-connected clusters that would have been removed in the
global algorithm due to the boundary constraint. In fact we have removed
\emph{all of them} for the following reason. Crucially any $2^{p}$-connected
clusters fit into a $B_{j\alpha}$, which will fit comfortably into
one of the boxes comprising a $P_{ab}$, with the distance between
$B_{j\alpha}$ and the boundary of the box being at least $2^{p_{*}}$.

So, for example, if a neutral $2^{p}$ connected cluster $C$ is sufficiently
within an element $P_{00}$, it will be dealt with identically to
the global algorithm. If it is close to, or straddles the boundary
of a box in$P_{00}$, it must be left alone at the first stage. However,
we do know that $C$ will lie entirely and deeply inside---at least
distance $2^{p_{*}}$ from the boundary---of a square in one of $P_{01,10,11}$.
Thus $C$ will be removed in one of these three subsequent steps.
Therefore, by then end of the level-$p$ round, we can be sure that
$C$ is transformed in the same way that it would be if we were to
use the global algorithm. 

\end{document}